# Spin-Polarized Magnetic Metal Electrodes for Magnetic Tunnel Junctions

*Zhiyuan Duan, Peixin Qin*, Li Liu, Guojian Zhao, Sixu Jiang, Xiaoyang Tan, Jingyu Li, Xiaoning Wang, Ziang Meng*, Zhiqi Liu**

Z. Duan, P. Qin, L. Liu, G. Zhao, S. Jiang, X. Tan, J. Li, Z. Meng, Z. Liu

School of Materials Science and Engineering, Beihang University, Beijing, 100191, China.

E-mail: qinpeixin@buaa.edu.cn; mengza@buaa.edu.cn; zhiqi@buaa.edu.cn

Z. Duan, P. Qin, L. Liu, G. Zhao, S. Jiang, X. Tan, J. Li, Z. Meng, Z. Liu

State Key Laboratory of Tropic Ocean Engineering Materials and Materials Evaluation, Beihang University, Beijing, 100191, China.

X. Wang

The Analysis & Testing Center, Beihang University, Beijing, 100191, China.

**Funding:**

Beijing Natural Science Foundation (JQ23005)

National Natural Science Foundation of China (52401300, 524B2003, 525B2008, 52425106, 52121001, 52271235, U25A20244)

National Key R&D Program of China (2022YFA1602700)

Fundamental Research Funds for the Central Universities

**Abstract**

Magnetic tunnel junctions are foundational components of spintronic memory, sensing, and computing, and their performance depends critically on the magnetic metallic electrodes. This Perspective examines electrode materials through the lens of magnetic order and the distinct microscopic mechanisms that generate spin-selective tunneling. Conventional ferromagnets, including CoFeB and half-metallic Heusler alloys, support exchange-split electronic states and symmetry-filtered tunneling, whereas compensated collinear and noncollinear antiferromagnets exploit sublattice selectivity, spin–orbit anisotropy, vector spin textures, and magnetic multipoles. Altermagnets provide a collinear, zero-net-moment route based on symmetry-allowed momentum-dependent spin splitting. Across these material classes, we compare the origins of tunneling polarization, the roles of barrier evanescent states and interface termination, and strategies for electrically writing and reading the relevant magnetic order. This comparison reveals a broader design principle: effective electrode polarization is not a scalar bulk quantity, but a momentum-, orbital-, symmetry-, and interface-resolved transport property. Beyond the pursuit of ever larger tunneling magnetoresistance, future progress will depend on converting the unconventional spin polarization of emerging magnetic metals into electrically addressable, thermally robust, and reproducible tunneling functionality at realistic interfaces.

## 1. Introduction

The magnetic tunnel junction (MTJ), consisting of two magnetic electrodes separated by an ultrathin insulating barrier, has evolved from a fundamental quantum-tunneling structure into a cornerstone of modern spintronic technologies, enabling magnetoresistive random-access memory (MRAM), magnetic sensors, and emerging logic and neuromorphic computing architectures [1-8]. Its key functional response, tunneling magnetoresistance (TMR), originates from the dependence of tunneling conductance on the magnetic states of the electrodes and is therefore governed by the spin selectivity of the electronic states transmitted across the barrier [9-11]. The development of MTJs can consequently be viewed not only as a history of improving barrier quality and device fabrication, but also as a continuing search for magnetic metallic electrodes capable of generating stronger, more controllable, and more robust spin-dependent tunneling.

The materials landscape of MTJ electrodes has expanded substantially during this development. Early devices relied on elemental ferromagnetic metals and transition-metal alloys, in which exchange splitting produces an imbalance between majority- and minority-spin states. The introduction of CoFe and CoFeB electrodes, together with crystalline MgO barriers, shifted the design strategy from increasing the density-of-states polarization alone to selecting electronic states with favorable momentum and orbital symmetry [10-14]. More recently, compensated magnetic metals have broadened the electrode concept beyond materials with a finite magnetization. Collinear antiferromagnets can generate tunneling responses through spin–orbit anisotropy or sublattice-selective interfaces, whereas noncollinear antiferromagnets exploit momentum-dependent vector spin textures and magnetic multipoles to support spin-polarized tunneling and electrical switching [15-18]. Altermagnetic metals provide a further collinear route, in which crystal symmetry permits momentum-dependent spin splitting despite vanishing net magnetization [19-23]. This materials evolution shows that a useful MTJ electrode need not possess a large macroscopic magnetization or even a finite Brillouin-zone-averaged spin polarization.

In this Perspective, we examine the evolution of spin-polarized magnetic metallic electrodes through the combined lenses of magnetic order, electronic structure, and interfacial transmission. We begin with the superconducting-tunneling experiments that established tunneling spectroscopy and spin-polarization measurements, and then discuss conventional ferromagnetic metals, half-metallic Heusler alloys, collinear and noncollinear antiferromagnetic metals, and emerging altermagnetic electrodes (Figure 1). Across these material classes, we focus on how spin-selective tunneling is generated, how the relevant magnetic order is written and read, and how electrode states couple to barrier evanescent states and realistic interfaces. A central theme is that effective electrode polarization is not simply a scalar bulk quantity, but a momentum-, orbital-, symmetry-, and interface-resolved transport property. This broader viewpoint provides a unified basis for evaluating established and emerging electrode materials and for developing next-generation MTJs with large electrical readout, efficient magnetic control, and reduced stray fields.

## 2. Historical Development of Tunneling and Spin-Dependent Transport

The story of magnetic tunneling begins not with magnetism, but with superconductivity. In 1960, Giaever demonstrated that electrons could tunnel through a thin insulating barrier separating a superconductor and a normal metal, revealing current–voltage characteristics that directly reflected the superconducting density of states [24]. The experimental concept was remarkably simple: a thin aluminum film was oxidized in air to form a native $Al_2O_3$ tunnel barrier and subsequently contacted by a second metallic electrode. Despite this simplicity, the experiment established a fundamental principle that would shape the future of tunneling research: quantum tunneling through an insulating barrier provides a powerful spectroscopic approach for probing the electronic states of electrodes. Two years later, Josephson predicted that Cooper pairs could tunnel coherently between two superconductors separated by a thin insulating barrier (Figure 2a), generating a dissipationless supercurrent governed by the phase difference between their macroscopic quantum states [25]. Experimental evidence for this effect was reported shortly thereafter by Anderson and Rowell, who observed the characteristic zero-voltage current and magnetic-field-dependent interference behavior expected for Josephson tunneling [26]. In parallel, Ambegaokar and Baratoff developed the microscopic theory of tunneling between superconductors and derived the quantitative relationship between the Josephson critical current, the normal-state junction resistance, and the superconducting energy gap [27]. Together, these studies transformed the Josephson effect from a theoretical prediction into a quantitatively established tunneling phenomenon. For the subsequent development of magnetic tunnel junctions, the superconducting tunneling era established several foundational concepts: the electrode/barrier/electrode architecture, the use of tunneling conductance as a spectroscopic probe of electrode electronic states, the preservation of quantum coherence across an insulating barrier, and the decisive role of the electrode wave functions and their coupling through the barrier in determining transport.

Having established that tunneling could probe the quasiparticle excitation spectrum and preserve the phase coherence of a superconducting condensate, the next natural question was whether it could also resolve the electron spin degree of freedom. A crucial step toward this goal was the observation by Meservey *et al.* that an intense magnetic field applied parallel to an ultrathin superconducting Al film Zeeman-splits its quasiparticle density of states into spin-resolved subbands [28]. Tedrow and Meservey subsequently exploited this spin-split superconducting density of states as a spectroscopic spin analyzer, first demonstrating spin-dependent tunneling between superconducting Al and ferromagnetic Ni and then extending the technique to Fe, Co, Ni, and Gd [29-30] . In a superconductor/insulator/ferromagnet junction, the Zeeman-split quasiparticle density of states produces four characteristic conductance features in the differential-conductance spectrum. Because majority- and minority-spin electrons in the ferromagnetic electrode contribute differently to the tunneling current, the corresponding conductance peaks exhibit unequal amplitudes. Quantitative analysis of this spectral asymmetry, with spin–orbit scattering and orbital pair-breaking effects taken into account, provides a direct measure of the effective tunneling spin polarization, $P_T = (G\uparrow - G\downarrow)/(G\uparrow + G\downarrow)$, where $G\uparrow$ and $G\downarrow$ denote the spin-up and spin-down contributions to the tunneling conductance, respectively [30-31]. The original measurements yielded tunneling spin polarizations of approximately 44% for Fe, 34% for Co, 11% for Ni, and 4.3% for Gd, demonstrating that majority-spin carriers dominate the tunneling current in these elemental ferromagnets [30].

More importantly, these experiments established that the polarization measured in tunneling cannot generally be identified with the unweighted bulk density-of-states polarization $P_{DOS} = [N\uparrow(E_F) - N\downarrow(E_F)]/[N\uparrow(E_F) + N\downarrow(E_F)]$, where $N\uparrow(E_F)$ and $N\downarrow(E_F)$ represent the spin-resolved density of states at the Fermi level. Instead, $P_T$ represents an effective transport polarization determined by the spin-dependent tunneling matrix elements, which describe the coupling between electronic states in the electrode and the barrier, as well as by orbital symmetry, interfacial electronic structure, and the barrier-dependent decay of electronic wave functions [1, 31-32]. Although the full implications of this distinction were not yet apparent, it foreshadowed a central principle of crystalline MTJs: tunneling currents are governed not simply by the total number of spin-polarized states at the Fermi level, but by the symmetry, crystal momentum, and barrier transmission probability of the specific Bloch states that couple efficiently through the barrier [1, 12, 33].

The transition from measuring spin polarization to exploiting it as a functional transport property was achieved by Jullière in 1975, who fabricated a Fe/Ge/Co tunnel junction and observed a 14% resistance change between the parallel and antiparallel magnetization configurations at 4.2 K (Figure 2b) [9]. Jullière introduced the phenomenological relation: TMR $= 2P_1P_2/(1-P_1P_2)$, where $P_1$ and $P_2$ represent the effective spin polarizations of the two magnetic electrodes, respectively. This model established the first quantitative connection between electrode spin polarization and tunnel magnetoresistance, transforming spin polarization from a spectroscopic observable into a device-relevant design parameter for magnetic tunnel junctions. Within the Jullière framework, the barrier is treated as a spin-independent and structureless tunneling medium, while electrode polarization is described as a scalar quantity averaged over all electronic states participating in transport. Subsequent theoretical developments, notably the work of Slonczewski, established a more rigorous quantum-mechanical description of tunneling conductance and interlayer exchange coupling between ferromagnetic electrodes separated by an insulating barrier [34]. This framework explicitly incorporated the electronic wave functions, tunneling matrix elements, and spin-dependent transmission probabilities that govern transport across the barrier. Despite these simplifications, the model provided a powerful guiding principle for several decades of MTJ development: increasing the spin polarization of magnetic electrodes is the most direct route toward enhancing TMR. It also predicted an ideal limit in which two half-metallic electrodes with $P=1$ would generate infinitely large TMR, motivating extensive efforts to identify and engineer materials with nearly complete spin polarization. However, the scalar description of polarization in the Jullière model also represents its fundamental limitation, because it neglects the roles of crystal symmetry, orbital selectivity, and conservation of transverse crystal momentum in determining which electronic states actually dominate tunneling transport.

Although Jullière's pioneering work did not immediately translate into a widely adopted technology, a parallel development in metallic multilayers established spin-dependent transport as a practical device concept and accelerated the emergence of spintronics. Building on earlier observations of antiferromagnetic interlayer exchange coupling between ferromagnetic layers separated by ultrathin Cr spacers [35], Fert and Grünberg independently discovered giant magnetoresistance (GMR) in Fe/Cr and Fe/Cr/Fe multilayer structures in the late 1980s (Figure 2c,d) [36-37]. Unlike TMR, where electrons traverse an insulating barrier through spin-dependent tunneling, GMR originates from spin-dependent scattering of conduction electrons at interfaces and within metallic layers. In the framework of the two-current model, spin-up and spin-down electrons experience different scattering rates, leading to a resistance change when the relative magnetic configuration of the ferromagnetic layers is modified. Although the microscopic mechanisms differ, GMR and TMR share the same fundamental principle: electrical transport in magnetic heterostructures is governed by the spin-dependent transmission of charge carriers.

Subsequent work by Parkin and co-workers established sputtered metallic multilayers as a practical platform for GMR (Figure 2e,f), demonstrating large and highly reproducible room-temperature magnetoresistance, oscillatory interlayer exchange coupling as a function of spacer thickness, and the spin-valve architecture based on a magnetically soft free layer and an exchange-biased reference layer separated by a nonmagnetic metallic spacer [38-41]. These advances established the materials growth, multilayer engineering, and magnetic-state control required for artificial magnetic heterostructures and enabled the commercialization of GMR-based read heads in hard-disk drives. The technological impact of GMR was further recognized by the 2007 Nobel Prize in Physics awarded to Fert and Grünberg. From the perspective of MTJ development, GMR played two critical roles. First, it validated spin-dependent transport as a technologically viable paradigm and established the importance of controlling magnetic configurations in nanoscale heterostructures. Second, the relatively limited resistance contrast achievable in metallic spin valves motivated the search for alternative architectures. This demand naturally revived interest in tunneling-based structures, where magnetoresistance arises from spin-dependent transmission through an insulating barrier rather than scattering-limited transport in metallic systems.

The technological validation of spin-dependent transport by GMR, together with the demand for a larger electrical readout signal than that available from metallic spin valves, set the stage for the revival of tunneling magnetoresistance. In 1995, two independent studies demonstrated sizable TMR at room temperature using amorphous $Al_2O_3$ barriers. Moodera and co-workers reported resistance changes of 11.8% at 295 K (Figure 3a), 20% at 77 K, and 24% at 4.2 K in CoFe/$Al_2O_3$/Co or NiFe junctions [10]. In parallel, Miyazaki and Tezuka observed comparable TMR ratios of 18% at 300 K and 30% at 4.2 K in Fe/$Al_2O_3$/Fe junctions (Figure 3b) [11]. These results established that spin-dependent tunneling could be sustained at technologically relevant temperatures and transformed the MTJ from a low-temperature proof of concept into a viable device architecture.

The subsequent development of $Al_2O_3$-based MTJs focused on increasing the effective interfacial spin polarization while simultaneously improving barrier uniformity, oxidation control, and thermal stability. A representative advance was reported by Han and co-workers, who obtained TMR ratios of 49.7% at room temperature and 69.1% at 4.2 K in annealed $Co_{75}Fe_{25}/AlO_x/Co_{75}Fe_{25}$ junctions (Figure 3c) [42-43]. This work demonstrated that electrode composition and post-growth annealing could strongly modify the effective barrier height, interfacial electronic structure, and measured tunneling polarization. Further optimization, including the introduction of amorphous CoFeB electrodes with smooth $Al_2O_3$ interfaces, raised the room-temperature TMR ratio to approximately 70% by 2004 [44]. These advances established amorphous $AlO_x$ barriers as a robust technological platform and underscored that TMR depends not only on the bulk polarization of the electrodes but also on the microscopic quality of both electrode/barrier interfaces.

Nevertheless, amorphous $AlO_x$ barriers imposed a practical ceiling on further TMR enhancement because they did not provide a well-defined crystalline symmetry filter. In the absence of long-range translational order, transverse crystal momentum $k_{\|}$ is not a strictly conserved quantum number across the barrier, and tunneling samples a broad ensemble of electronic states with different orbital characters and attenuation rates. The resulting conductance is therefore an effective average over many transmission channels. Highly spin-polarized states may contribute only weakly if their wave functions couple inefficiently to the barrier, whereas less strongly polarized states with larger tunneling matrix elements may dominate transport. Thus, the TMR of an amorphous junction is governed by a barrier- and interface-weighted transport polarization rather than by the unweighted bulk density-of-states polarization alone [1]. The approximately 70% room-temperature TMR reached in optimized $Al_2O_3$ junctions consequently represented not a failure of the Jullière principle, but the limit of treating all accessible tunneling states without strong momentum or symmetry selection.

The route beyond this limit was therefore not simply to identify electrodes with a larger bulk spin polarization, but to engineer a barrier that selectively transmits the most strongly spin-polarized electronic states. In 2001, Butler *et al.* and Mathon and Umerski independently predicted that crystalline MgO could act as such a symmetry filter in epitaxial Fe/MgO/Fe(001) junctions [12, 33]. Their first-principles transport calculations showed that electronic states of different symmetry decay at markedly different rates within the MgO band gap. Along the [001] tunneling direction, the $\Delta_1$ evanescent state, derived primarily from $s$, $p_z$, and $d_{z^2}$ orbitals, exhibits the slowest attenuation and therefore dominates transport through sufficiently thick MgO barriers. In bcc Fe and CoFe electrodes, a majority-spin $\Delta_1$ band crosses the Fermi level near the Brillouin-zone center, whereas a corresponding minority-spin $\Delta_1$ state is absent at $E_F$. Consequently, the parallel magnetic configuration supports efficient majority-spin $\Delta_1$ transmission through both electrodes, while this channel is strongly suppressed in the antiparallel configuration. MgO therefore enhances TMR not by transmitting all majority-spin electrons equally, but by preferentially selecting Bloch states with the appropriate crystal momentum and orbital symmetry. The ideal calculations consequently predicted TMR ratios of the order of 1000%, far beyond the values achievable with amorphous barriers [12, 33].

Experimental confirmation followed in 2004. Yuasa *et al.* demonstrated 180% room-temperature TMR in fully epitaxial Fe/MgO/Fe junctions grown by molecular beam epitaxy (Figure 4a) [13]. In the same year, Parkin *et al.* reported approximately 220% at room temperature in sputter-deposited, highly (001)-textured CoFe/MgO/CoFe junctions (Figure 4b,c) [14]. The latter result was particularly important because it showed that strong symmetry-selective tunneling could be realized using scalable sputtering processes rather than being restricted to ideal molecular-beam-epitaxy structures. CoFeB/MgO/CoFeB junctions followed shortly thereafter, reaching 230% room-temperature TMR in 2005 [45]; annealing-induced crystallization of initially amorphous CoFeB electrodes subsequently enabled 604% at 300 K in 2008 [46]. More recently, precision control of the crystallographic orientation and MgO interfaces in epitaxial CoFe/MgO/CoFe(001) junctions increased the room-temperature TMR record to 631%, together with 1143% at 10 K (Figure 4d,e) [47]. This result is especially revealing because the record was obtained not by introducing a nominally more highly polarized bulk electrode, but by optimizing atomic-scale interface structure and the transmission of the $\Delta_1$ channel. The progression from approximately 70% in amorphous $Al_2O_3$ junctions, through 604% in CoFeB/MgO/CoFeB, to 631% in epitaxial CoFe/MgO/CoFe therefore reinforces the central principle of this Perspective: the decisive electrode property is not the Brillouin-zone-averaged spin polarization, but the momentum-resolved and symmetry-projected polarization of the states that couple efficiently across the electrode/barrier interface.

This conceptual shift fundamentally broadened the criteria for identifying spin-polarized electrodes. Once tunneling is understood as a momentum- and symmetry-selective process, a large net magnetization is no longer the only route to a spin-asymmetric conductance. Any magnetic material that provides strongly unequal transmission probabilities for the relevant electronic states can, in principle, function as an effective MTJ electrode. This insight provides the natural starting point for considering ferromagnets, collinear/noncollinear antiferromagnets, and altermagnets in the sections that follow.

**3. Ferromagnetic Metallic Electrodes**

Within the symmetry-selective framework established above, ferromagnetic metals remain the benchmark electrodes for MTJs. Exchange splitting produces an imbalance between the electronic states available to the two spin channels, while the finite magnetization provides a readily controllable order parameter for defining parallel and antiparallel resistance states. Elemental Fe, Co, and Ni established the basic principle of spin-polarized tunneling, but their moderate and strongly material-dependent tunneling polarizations limited the attainable TMR. Alloying Fe with Co increases the effective transport polarization and, more importantly, provides a bcc electronic structure that couples favorably to the slowly decaying $\Delta_1$ evanescent states of crystalline MgO. These advantages established CoFe and its boron-containing derivatives as the dominant family of conventional ferromagnetic metallic electrodes.

Among these materials, CoFeB provides the clearest illustration of why the performance of an MTJ electrode cannot be predicted from its bulk spin polarization alone. As deposited, CoFeB is an amorphous metallic alloy that forms smooth and chemically uniform interfaces with ultrathin MgO barriers. During post-deposition annealing, crystallization proceeds rapidly from the CoFeB/MgO interface, producing a bcc-CoFe(001)-like structure templated by the MgO barrier [45-46, 48-49]. At the same time, B is redistributed away from the developing crystalline CoFe region. Its eventual location depends strongly on the adjacent seed and capping layers: B may be absorbed by metallic layers such as Ta, remain segregated near an interface, or enter the MgO barrier and form B-containing interfacial species [50-52]. This chemical redistribution is critical because residual B at the CoFeB/MgO interface can suppress the majority-spin $\Delta_1$ conductance and reduce TMR. The technological success of CoFeB therefore arises not from an intrinsically exceptional bulk polarization, but from a processing synergy that combines amorphous-state interface smoothness, annealing-induced bcc crystallization, controlled B redistribution, and coherent $\Delta_1$-symmetry tunneling.

This processing synergy enabled room-temperature TMR ratios of 230% in CoFeB/MgO/CoFeB junctions in 2005, followed by values above 300% in thermally stable synthetic-pinned-layer structures and 604% after optimization of high-temperature annealing and suppression of detrimental Ta diffusion [45-46, 48]. The same metal/oxide interface also generates perpendicular magnetic anisotropy in sufficiently thin CoFeB layers. First-principles calculations attribute this anisotropy to spin–orbit-coupling-induced changes in the hybridized Fe(Co) 3*d* and O 2*p* states near the Fermi level, with the magnitude being highly sensitive to interfacial oxidation and atomic coordination [53-54]. The coexistence of large TMR, interfacial perpendicular anisotropy, sputter-process compatibility, and adequate thermal stability made CoFeB/MgO/CoFeB the dominant materials platform for perpendicular spin-transfer-torque MRAM. Its success demonstrates that the electrode, barrier, adjacent layers, and annealing process must be designed as an integrated interfacial system rather than optimized independently.

A complementary strategy is to employ half-metallic ferromagnetic metals. In an ideal half-metal, one spin channel is metallic whereas the other possesses an electronic band gap at the Fermi level, leading to a nominal spin polarization of 100% [55]. Co-based full-Heusler alloys, particularly $Co_2MnSi$ and compositionally tuned $Co_2(Mn,Fe)Si$, have received sustained attention because they combine a high Curie temperature with a theoretically predicted minority-spin gap and a crystal structure compatible with epitaxial MgO barriers. Their metallic majority-spin channel supplies the tunneling current, while the absence of minority-spin states at $E_F$ ideally suppresses the opposite-spin conductance. The performance of $Co_2MnSi$ electrodes is strongly dependent on composition and chemical ordering. Systematic studies of $Co_2MnSi$/MgO/$Co_2MnSi$ junctions showed that Mn-rich compositions produce substantially higher TMR than Mn-deficient compositions [56]. Following this optimization, fully epitaxial $Co_2MnSi$/MgO/$Co_2MnSi$ junctions achieved TMR ratios of 1995% at 4.2 K and 354% at 290 K (Figure 4f–h), demonstrating the combined action of an intrinsically highly spin-polarized electrode and coherent tunneling through MgO [57]. Independent spin-resolved photoemission measurements observed a polarization of approximately 90–93% at the Fermi level in the near-surface region of well-ordered $Co_2MnSi$ films at room temperature [58]. More recent bulk-sensitive spin-resolved hard-X-ray photoemission further revealed a metallic majority-spin Fermi edge and a minority-spin gap in MgO-capped $Co_2MnSi$, with the half-metallic character persisting up to room temperature [59]. These spectroscopic results provide complementary evidence that a highly spin-polarized metallic electronic structure can be retained in properly ordered thin films and beneath a protective oxide layer.

Nevertheless, Heusler electrodes also reveal the fragility of ideal bulk half-metallicity at a real tunnel interface. Co antisite defects in Mn-deficient $Co_2MnSi$ introduce electronic states into the nominal minority-spin gap, whereas Mn-rich compositions suppress detrimental Co antisites and more effectively preserve the half-metallic electronic structure [56, 60]. The actual tunneling polarization is additionally affected by the degree of $L2_1$ or B2 chemical order, atomic termination, interdiffusion, strain, interface resonances, and thermally activated spin-flip channels. This sensitivity explains why the TMR decreases strongly with increasing temperature even though the Curie temperature of $Co_2MnSi$ remains far above room temperature. Spin- and momentum-resolved photoemission has further shown that the minority-spin gap and the momentum-dependent polarization texture $P(E, k)$ can be tuned through band filling and alloy composition [61]. Thus, the relevant quantity is not simply whether the ideal bulk compound is classified as a half-metal, but whether the specific states selected by the interface and barrier retain their spin asymmetry under realistic structural, chemical, and thermal conditions.

CoFeB and Co-based Heusler alloys therefore represent two complementary design philosophies for ferromagnetic metallic electrodes. CoFeB derives its performance primarily from processing compatibility, interface crystallization, and barrier-induced symmetry filtering, whereas Heusler alloys seek to combine coherent symmetry filtering with intrinsically high bulk spin polarization. The comparison between these two families establishes a general design rule that extends to all magnetic-metal electrode classes discussed below: an effective MTJ electrode must simultaneously provide a metallic conduction channel, a strong spin asymmetry in the states selected for tunneling, appropriate momentum and orbital matching to the barrier, an atomically stable interface termination, and magnetic properties compatible with writing and retention. Bulk spin polarization alone is neither a sufficient predictor of TMR nor a sufficient criterion for selecting an electrode material.

Despite their remarkable success, ferromagnetic metallic electrodes retain intrinsic scaling trade-offs that become increasingly important as devices are pushed toward higher density and shorter latency. Their finite magnetization produces dipolar stray fields that can bias the free layer, distort switching symmetry, and increase magnetic cross-talk and write-error variability. Far from a finite magnetic element, the dipolar field decreases approximately as $r^{-3}$, although its magnitude and spatial distribution depend strongly on the device geometry and magnetic configuration. Synthetic antiferromagnetic reference layers, in which two ferromagnetic layers are coupled antiparallel across a nonmagnetic spacer, are therefore widely used to reduce the net magnetic moment and compensate the stray field. This approach substantially mitigates the problem, but it introduces additional layers and processing complexity and may leave residual fields when the magnetic moments of the constituent layers are not perfectly balanced [62-63]. Current-induced switching by spin-transfer torque or spin-orbit torque also requires charge current and consequently generates Joule heating. The switching current is coupled to thermal stability because both depend on magnetic volume and anisotropy, but it is not determined by the thermal stability factor $\Delta = E_b/k_B T$ alone. Magnetic damping, spin-torque efficiency, effective magnetic fields, pulse duration, and thermal activation also play important roles [64-65]. These coupled requirements create a retention, write-energy, and read-disturb trade-off in scaled MTJs. Ferromagnetic resonance and magnetization precession commonly occur in the gigahertz regime, but they do not impose a universal switching-time limit of approximately 1 ns. Optimized spin-orbit-torque devices have demonstrated subnanosecond and even picosecond magnetization reversal [66-67]. Nevertheless, zero-net-moment magnetic metals remain attractive because they can strongly suppress dipolar stray fields and support exchange-enhanced magnetic dynamics. These advantages provide a fundamental motivation for exploring compensated magnetic metals as active electrodes in magnetic tunnel junctions.

## 4. Collinear Antiferromagnetic Metallic Electrodes

Conventional collinear antiferromagnetic metals provide the most direct extension of the MTJ electrode concept from ferromagnets with a finite magnetization to compensated magnetic order [18, 68-78]. Their oppositely aligned magnetic sublattices produce a vanishing or nearly vanishing macroscopic magnetization, while the Néel vector describes the orientation of the staggered order. In many conventional collinear antiferromagnets, crystal symmetries enforce spin-degenerate bulk bands or an exact cancellation of the spin polarization carried by the two magnetic sublattices. The scalar polarization entering the Jullière model therefore vanishes, apparently precluding conventional TMR. Early theoretical studies nevertheless showed that antiferromagnetic metals can generate magnetoresistance and spin-transfer torque when coherent transport preserves the sublattice-resolved spin information across the junction [79]. This concept was subsequently formulated in terms of Néel spin currents, in which currents associated with different magnetic sublattices carry opposite spin polarizations despite a vanishing net spin polarization. If the interfaces preserve or selectively probe this sublattice character, the staggered current can generate TMR and Néel-vector torque [80]. This conventional spin-compensated class should be distinguished from altermagnetic metals, whose bulk electronic bands exhibit symmetry-allowed momentum-dependent spin splitting.

CuMnAs and $Mn_2Au$ represent two prototypical metallic collinear antiferromagnets considered as active electrodes in theoretical MTJ studies. For CuMnAs/GaP/CuMnAs junctions, Stamenova *et al.* predicted a conductance contrast between different relative orientations of the two Néel vectors, together with a staggered spin-transfer torque acting oppositely on the magnetic sublattices [81]. The torque retained its staggered character for different interface terminations, whereas the magnitude and sign of the magnetoresistance were sensitive to the atomic structure of the CuMnAs/GaP interfaces. Related calculations for Nb/$Mn_2Au$/CdO/$Mn_2Au$/Nb junctions predicted TMR ratios of the order of 1000% for selected ideal structures [82]. The large calculated response was attributed to momentum-selective tunneling through CdO and interfacial resonant states at the $Mn_2Au$/CdO boundaries, but it was strongly affected by interface termination, atomic vacancies, intermixing, and applied bias. Although these predictions have not yet been experimentally verified, they demonstrate that spin-degenerate collinear antiferromagnetic metals can, in principle, support substantial TMR when carefully designed barriers and interfaces resolve their staggered magnetic order.

The first experimental implementation of an antiferromagnetic metal in a tunnel junction relied on tunneling anisotropic magnetoresistance (TAMR) rather than the parallel–antiparallel response of a conventional MTJ. Unlike conventional tunneling magnetoresistance (TMR), which originates from the relative alignment of spin-polarized states in two magnetic electrodes, TAMR arises from the dependence of the tunneling conductance on the orientation of a single magnetic electrode's order parameter. In particular, spin–orbit coupling modifies the electronic structure and tunneling matrix elements when the Néel vector rotates relative to the crystal axes, resulting in an orientation-dependent resistance. Park *et al.* reported a spin-valve-like resistance change exceeding 100% in NiFe/IrMn/MgO/Pt junctions in 2011 (Figure 5a,b) [83]. The NiFe layer acted as an exchange-spring actuator for the metallic IrMn electrode, while the measured resistance change arose from the orientation-dependent tunneling electronic structure of IrMn. Room-temperature operation was subsequently achieved in [Pt/Co]/IrMn/$AlO_x$/Pt junctions [84]. Ferromagnet-free Ta/MgO/IrMn junctions further demonstrated resistance changes of up to approximately 10% between states prepared by field cooling along different directions [85]. These studies established that the Néel order of a metallic antiferromagnetic electrode can directly control tunneling conductance, although the required exchange-spring or field-cooling procedures were not suitable for scalable electrical writing.

In 2019, Yan *et al.* demonstrated electric-field control of metallic $L1_0$-MnPt on a piezoelectric PMN–PT substrate, where strain-induced redistribution of the Néel-vector orientations produced nonvolatile room-temperature resistance states that remained stable in magnetic fields up to 60 T [86]. In a Pt/$MgAl_2O_4$/MnPt tunnel junction, the same mechanism yielded a room-temperature TAMR ratio of approximately 11.2% (Figure 5c,d), with MnPt acting directly as the magnetic metallic electrode. The response nevertheless remained a single-electrode TAMR rather than conventional TMR between two antiferromagnetic electrodes. Complementing this single-antiferromagnet-electrode approach, Du *et al.* demonstrated a Pt/IrMn/CoFeB/MgO/CoFeB three-terminal junction, in which current pulses as short as 0.8 ns switched the IrMn state and exchange coupling transferred this information to the adjacent

CoFeB layer, enabling a TMR readout exceeding 80% (Figure 5e,f) [87]. Because the large resistance contrast was generated by the conventional CoFeB/MgO/CoFeB tunneling channel rather than by direct spin-polarized tunneling from IrMn, this device represents a hybrid antiferromagnet–ferromagnet architecture rather than an all-antiferromagnetic MTJ.

Two-dimensional van der Waals antiferromagnets provide a complementary platform for antiferromagnetic tunneling because their layer number, stacking sequence, and interface termination can be controlled at nearly the atomic scale. Earlier CrSBr spin-filter junctions had already demonstrated exceptionally large field-dependent TMR, establishing A-type antiferromagnetic semiconductors as efficient spin-selective tunnel barriers [88]. In 2024, Chen *et al.* introduced a twist between two CrSBr bilayers and obtained more than 700% nonvolatile TMR at zero magnetic field (Figure 5g,h) [89]. The twist weakens the magnetic coupling at the central interface and stabilizes distinct relative spin configurations, while the strong twist-angle dependence of the conductance reflects coherent tunneling governed by the momentum-dependent decay of evanescent states. CrSBr is, however, an *n*-type antiferromagnetic semiconductor, and the CrSBr stack itself serves as the magnetic tunnel barrier rather than as a metallic electrode. A direct realization of an all-collinear-antiferromagnetic junction with magnetic metallic electrodes was subsequently achieved using the van der Waals A-type antiferromagnetic metal $(Fe_{0.6}Co_{0.4})_5GeTe_2$, abbreviated as FCGT, on both sides of a $WSe_2$ barrier [90-92]. In FCGT, the magnetic moments are aligned ferromagnetically within each van der Waals layer and antiferromagnetically between adjacent layers. The resulting junctions exhibited TMR ratios of up to 75% at 10 K within the antiferromagnetic phase (Figure 5i). The even–odd layer dependence produced either volatile or nonvolatile responses, demonstrating the decisive role of layer parity and interface termination. Transport calculations based on an idealized A-type $Fe_5GeTe_2$ electrode showed spin-degenerate bulk conduction channels but strongly spin-dependent transmission at the uncompensated interfaces. The effective tunneling polarization therefore originates from the magnetic sublattice exposed to the barrier rather than from a uniformly spin-polarized bulk band structure. This comparison illustrates two distinct roles of two-dimensional antiferromagnets in tunnel junctions. Semiconducting CrSBr acts as a spin-filtering magnetic barrier, whereas metallic FCGT functions as a genuine active electrode whose polarization is generated at the interface.

The development of collinear antiferromagnetic metallic electrodes reveals two principal mechanisms for tunneling readout. The first relies on spin–orbit-coupling-induced anisotropy, which makes the electronic structure and tunneling matrix elements sensitive to the orientation of a single Néel vector and thereby produces TAMR. The second relies on sublattice-selective transport or uncompensated interface termination, which converts the locally staggered magnetic order into spin-dependent transmission and can generate a genuine two-electrode TMR response. In both cases, the tunneling signal does not originate from a uniform bulk spin polarization. Instead, it emerges from the symmetry-dependent interfacial electronic structure or from atomic-scale selection of one component of the staggered magnetic order. Increasing the operating temperature, achieving deterministic zero-field electrical switching, controlling antiferromagnetic domains, and reproducing the required interface termination remain central challenges. Nevertheless, these mechanisms establish collinear antiferromagnetic metals as a viable extension of the spin-polarized electrode concept beyond materials with a net magnetization.

## 5. Noncollinear Antiferromagnetic Metallic Electrodes

Whereas conventional collinear antiferromagnetic metals generally require spin–orbit-induced anisotropy or sublattice-selective interfaces to overcome bulk spin compensation, noncollinear antiferromagnetic metals provide a distinct route to spin-dependent tunneling. The nonparallel arrangement of their local moments removes a global spin-quantization axis and can produce spin-split Bloch states with vector spin expectation values that vary across the Fermi surface. Although the spin polarization may cancel when integrated over the entire Brillouin zone, individual transverse-momentum channels can remain strongly spin polarized. A crystalline tunnel barrier can selectively transmit these channels, allowing the relative orientation of the noncollinear magnetic orders in the two electrodes to produce a substantial conductance contrast [93-95].

Representative metallic noncollinear antiferromagnets include the hexagonal $D0_{19}$-type kagome compounds $Mn_3Sn$, $Mn_3Ge$, and $Mn_3Ga$, together with the cubic $L1_2$-type compounds $Mn_3Pt$, $Mn_3Ir$, and $Mn_3Rh$ [96-107]. Their noncollinear magnetic order produces a vanishing or very small net magnetization but can lower the magnetic symmetry sufficiently to permit pronounced time-reversal-odd electronic responses. For symmetry-appropriate members of this family, the magnetic order can be classified using cluster magnetic multipoles. In particular, ferroically ordered cluster magnetic octupoles provide a unified order-parameter description of the anomalous Hall response in several hexagonal and cubic $Mn_3X$ compounds [108]. The large anomalous Hall responses observed or predicted across this materials family demonstrate that a strongly asymmetric momentum-space electronic structure does not require a correspondingly large macroscopic magnetization [96-97]. The anomalous Hall effect itself, however, is not a direct measurement of tunneling spin polarization. Its intrinsic contribution is governed by the Berry curvature of the occupied electronic bands, whereas TMR depends on the spin texture, transverse momentum, orbital symmetry, and transmission probability of the Fermi-level states selected by the barrier and interfaces.

The connection between noncollinear order and longitudinal spin transport was established theoretically by Železný *et al.*, who showed that charge currents in noncollinear antiferromagnetic metals can carry a finite spin polarization despite their nearly vanishing magnetization [93]. First-principles transport calculations subsequently predicted TMR ratios of up to approximately 300% in junctions with $Mn_3Sn$ electrodes (Figure 6a,b), together with four resistance states associated with different relative orientations of the two noncollinear orders [94]. More recently, an effective momentum-dependent polarization was introduced to quantify the vector spin texture of individual tunneling channels. Calculations revealed nearly complete spin polarization over substantial portions of the Fermi surfaces of several noncollinear antiferromagnetic metals. For $Mn_3GaN/SrTiO_3/Mn_3GaN$ junctions, matching these strongly polarized states to the slowly decaying evanescent states of $SrTiO_3$ was predicted to produce an extraordinary TMR approaching 10000% (Figure 6c,d) [95]. These results reinforce the principle that a small or vanishing total current polarization does not preclude a large TMR when the barrier selectively transmits highly polarized momentum-resolved channels.

Experimental realization of noncollinear all-antiferromagnetic tunnel junctions was achieved in 2023. Qin *et al.* constructed $Mn_3Pt/MgO/Mn_3Pt$ junctions in which a collinear MnPt layer exchange-biased one of the noncollinear $Mn_3Pt$ electrodes. The junctions exhibited nonvolatile room-temperature magnetoresistance approaching 100% (Figure 6e–g), and first-principles calculations attributed the conductance contrast to the momentum-dependent spin polarization of metallic $Mn_3Pt$ and its spin-dependent matching across the MgO barrier [109]. Because both the exchange-bias and tunneling layers were antiferromagnetic, this architecture provided a stable reference state and a large room-temperature readout without using a ferromagnetic electrode. In parallel, Chen *et al*. demonstrated an $Mn_3Sn/MgO/Mn_3Sn$ junction with a room-temperature TMR ratio of approximately 2% between parallel and antiparallel configurations of the cluster magnetic octupoles [110]. Measurements of $Fe/MgO/Mn_3Sn$ junctions further showed that the direction and sign of the longitudinal spin-polarized current were controlled by the octupole orientation. The all-antiferromagnetic TMR was substantially larger than that estimated from the small net current polarization using the scalar Jullière model, demonstrating the importance of momentum-dependent transmission associated with the noncollinear magnetic order.

Subsequent studies established fully electrical writing and reading in $Mn_3Pt$-based all-antiferromagnetic tunnel junctions. In 2024, Shi *et al*. fabricated sputter-deposited $Mn_3Pt/Al_2O_3/Mn_3Pt$ three-terminal junctions on thermally oxidized silicon, in which spin–orbit torque from an adjacent Pt layer electrically controlled the free antiferromagnetic electrode without an applied magnetic field [111]. The devices exhibited room-temperature TMR ratios of up to 110% (Figure 6h), with a resistance output more than three orders of magnitude larger than the corresponding anisotropic-magnetoresistance readout, while first-principles calculations attributed the effect to momentum-resolved spin-dependent tunneling between the two noncollinear metallic electrodes. This work extended the earlier epitaxial proof of principle to a scalable, silicon-compatible platform with separate write and read paths. More recently, Kang *et al.* reported octupole-driven spin-transfer-torque switching in nanoscale $Mn_3Pt/MgO/Mn_3Pt$ junctions fabricated on oxidized silicon [112]. In this two-terminal architecture (Figure 6i), the vertical tunneling current directly switched the cluster magnetic octupole and simultaneously provided the resistance readout, producing room-temperature TMR values of up to 363% at switching current densities of the order of 10 $MA/cm^2$ (Figure 6j). The proposed torque originates from an imbalance between intra-sublattice and inter-sublattice tunneling currents, which generates a finite staggered torque on the three magnetic sublattices despite the globally spin-neutral total current. Together, these studies demonstrate two complementary electrical writing strategies for noncollinear all-antiferromagnetic junctions, namely lateral spin–orbit-torque control in a three-terminal geometry and vertical octupole-driven spin-transfer torque in a two-terminal geometry.

The development of noncollinear antiferromagnetic metallic electrodes therefore expands electrode polarization from a scalar spin imbalance to a momentum-dependent vector or multipolar quantity. Their TMR does not require a globally spin-polarized density of states, but instead originates from the matching of spin-textured and sublattice-resolved conduction channels selected by the barrier. The progression from magnetic-field-controlled TMR to lateral spin–orbit-torque writing and vertical octupole-driven torque demonstrates that noncollinear antiferromagnetic order can support both electrical readout and current-induced switching. The magnitude and sign of these effects remain sensitive to magnetic symmetry, junction orientation, barrier electronic structure, interface termination, crystallographic texture, and antiferromagnetic domain configuration. Reducing the switching-current density, improving thermal stability and endurance, and achieving reproducible nanoscale interfaces remain important requirements for device applications.

**6. Altermagnetic Metallic Electrodes**

Altermagnetic metals provide a distinct collinear route to spin-polarized tunneling without a net magnetization [19-21, 113-115]. In many conventional collinear antiferromagnets, opposite-spin sublattices are related by a lattice translation or inversion operation, which protects spin-degenerate electronic bands in the nonrelativistic limit. In an altermagnet, the opposite-spin sublattices are instead connected by crystal rotations or mirror operations. The resulting spin-group symmetry permits an exchange-driven spin splitting that does not require spin–orbit coupling. This splitting has an alternating even-parity structure in momentum space, changing sign between crystal-symmetry-related regions while vanishing along symmetry-protected nodal lines or planes [22-23, 116-118]. Depending on the crystal and magnetic symmetries, the corresponding momentum-space spin order can exhibit *d*-, *g*-, or *i*-wave character. Altermagnets therefore retain the collinear spin axis and compensated real-space magnetic order of an antiferromagnet, while supporting spin-split Fermi-surface sectors analogous to those of a ferromagnet [20, 113].

For the present focus on magnetic metallic electrodes, CrSb, $RuO_2$, and $KV_2Se_2O$ are particularly relevant [119-126]. Although $RuO_2$ is an oxide and $KV_2Se_2O$ is an oxychalcogenide, both possess metallic electronic structures and therefore fall within the scope of magnetic metallic electrodes considered here. CrSb combines collinear compensated magnetic order with a Néel temperature well above room temperature and has emerged as one of the most firmly established metallic altermagnets. Photoemission measurements on epitaxial CrSb films revealed an exchange-driven band splitting of approximately 0.6 eV near the Fermi level [120], while subsequent three-dimensional and spin-resolved measurements established its bulk *g*-wave character, identified spin splitting approaching 1 eV, and directly resolved the spin polarization of the split bands [121]. Although the intrinsic magnetic ground state of bulk $RuO_2$ remains controversial, several thin-film experiments have reported transport and spectroscopic signatures consistent with altermagnetic order [127-140]. These results suggest that the proposed magnetic state may be stabilized under particular conditions of strain, crystallographic orientation, stoichiometry, and interfacial coupling, making thin-film $RuO_2$ a particularly relevant, albeit still unsettled, platform for MTJ studies. $KV_2Se_2O$ provides a complementary example: experiments have identified it as a room-temperature metallic *d*-wave altermagnet with a strongly anisotropic spin-polarized Fermi surface [125], although its application as an MTJ electrode remains primarily at the theoretical stage.

The tunneling mechanism of an altermagnetic electrode is intrinsically momentum selective. Although the spin polarization integrated over the full Brillouin zone can vanish, individual transverse-momentum channels may be strongly polarized. Reversing the Néel vector reverses the spin character of these momentum-resolved channels. When the two electrodes have equivalent Néel-vector configurations, strongly transmitting channels with compatible momentum and spin character can overlap across the barrier. Reversing one electrode can suppress this overlap and produce a large resistance contrast. The resulting TMR cannot generally be described by a scalar Jullière polarization because it depends on the overlap of the full spin-resolved transmission distributions in $k_{||}$ space. The original theoretical formulation predicted magnetoresistance on the scale of approximately 100% in representative unconventional collinear antiferromagnets (Figure 7a) [19]. Subsequent first-principles calculations for $RuO_2/TiO_2/RuO_2$ junctions predicted TMR of several hundred percent, governed by matching between momentum-dependent conduction channels rather than by the total transport polarization (Figure 7b–d) [141]. Calculations for hybrid $RuO_2/TiO_2/CrO_2$ junctions further predicted almost vanishing TMR along the [001] direction but values as high as approximately 6100% along [110] (Figure 7e,f), illustrating that crystallographic orientation is a fundamental design variable for altermagnetic electrodes [142].

$RuO_2$-based tunnel junctions have exhibited two distinct forms of Néel-order-dependent magnetoresistance. Noh *et al.* fabricated $RuO_2/TiO_2$/CoFeB junctions and observed a TMR signal of up to approximately 5% at 10 K (Figure 7g–i), whose sign reversed upon reversal of the $RuO_2$ Néel vector [143]. In this hybrid junction, the ferromagnetic CoFeB electrode acts as a spin analyzer, and the resistance change was attributed to the matching between its spin-polarized states and the anisotropic momentum-dependent spin polarization of the $RuO_2$ electrode. The use of a ferromagnetic counterelectrode and the disappearance of the signal above approximately 50 K distinguish this device from an all-altermagnetic MTJ. In $RuO_2$/MgO/$RuO_2$ junctions, Xu *et al.* reported a room-temperature resistance change of approximately 60% when a magnetic-field-induced spin-flop transition changed the Néel-vector configuration from parallel to approximately orthogonal (Figure 7j) [144]. Despite the use of two $RuO_2$ electrodes, the response was identified primarily as spin-flop tunneling anisotropic magnetoresistance rather than conventional interelectrode TMR. Calculations attributed the large signal mainly to the dependence of spin–orbit-coupled interfacial resonant states on the absolute Néel-vector orientation and to their transmission through the MgO barrier. These studies therefore demonstrate different routes by which the magnetic order of metallic $RuO_2$ films can modulate tunneling conductance: spin-dependent electrode matching in a hybrid $RuO_2$/ferromagnet junction and orientation-dependent interfacial transmission in an all-altermagnetic tunnel junction.

CrSb provides a complementary and experimentally well-established metallic platform for altermagnetic MTJs. Calculations for CrSb/$\beta$-InSe/CrSb junctions predicted a TMR ratio of approximately 290% at the calculated Fermi level and values above 850% following an energy shift (Figure 7k) [145]. More recent quantum-transport calculations showed that the TMR of CrSb electrodes is strongly anisotropic. Transport along symmetry-allowed directions was predicted to produce approximately 870% TMR, while matching the CrSb conduction channels to the low-decay evanescent states of MgO increased the calculated value to approximately 1700%. Barrier-thickness and carrier-density optimization produced still larger theoretical values [146]. First-principles calculations have further extended CrSb electrodes to multiferroic tunnel junctions with an $\alpha$-$In_2Se_3$ barrier, showing that interface termination and ferroelectric polarization can jointly control spin-channel matching and generate multiple nonvolatile resistance states [147]. These predictions have not yet been experimentally verified, but they are important because CrSb combines metallic conduction, a Néel temperature near 700 K, and experimentally established spin splitting close to the Fermi level.

$KV_2Se_2O$ provides a particularly instructive example of how Fermi-surface geometry, spacer electronic structure, barrier thickness, and interface termination jointly determine the performance of altermagnetic metallic electrodes. An all-metal $KV_2Se_2O/BaTi_2Bi_2O/KV_2Se_2O$ junction was predicted to exhibit tunnel-like transport because of suppressed electronic coupling through the metallic spacer, with symmetry-driven spin selectivity producing a calculated magnetoresistance ratio of $1.5\times10^{12}\%$ [148]. This architecture extends altermagnetic magnetoresistance beyond conventional junctions containing insulating tunnel barriers. The quasi-two-dimensional altermagnetic flat bands of $KV_2Se_2O$ were further shown to generate strongly anisotropic Fermi sheets, restricting the overlap between opposite-spin channels to a small number of nodal-like regions (Figure 7l) [149]. The resulting suppression of leakage transmission in the antiparallel configuration produced a calculated intrinsic TMR of approximately $4.3 \times 10^3\%$, which increased to approximately $1.1 \times 10^6\%$ upon introducing an insulating barrier. Calculations for $KV_2Se_2O/SrTiO_3/KV_2Se_2O$ junctions also predicted a pronounced oscillation of the tunneling conductance with the number of $SrTiO_3$ layers, arising from layer-parity-dependent interface configurations and effective barrier profiles; a four-layer $SrTiO_3$ barrier yielded a TMR of $4.6\times10^{7}$% [150]. The effects of experimentally relevant surface terminations were subsequently examined, with TMR values above $10^{5}$% predicted for all considered terminations. In particular, K-termination was found to preserve the bulk spin polarization through interfacial passivation, producing values as high as $10^{12}$% [151]. More recently, $KV_2Se_2O$/MgO/$KV_2Se_2O$ junctions were predicted to exhibit bias-induced negative differential resistance and TMR sign reversal due to the suppression of momentum-resolved tunneling channels [152]. Although these exceptionally large ratios correspond to ideal coherent-transport calculations, the results collectively demonstrate the strong potential of $KV_2Se_2O$ as a metallic electrode for high-performance altermagnetic junctions.

Taken together, these studies establish altermagnetic metals as a distinct class of spin-selective electrodes whose tunneling response originates from momentum-dependent spin splitting rather than from a net magnetization. Their magnetoresistance is governed by the combined effects of the intrinsic Fermi-surface geometry, crystallographic orientation, momentum-space channel matching, barrier or spacer electronic structure, and atomic-scale interface configuration. Although current results range from initial experimental demonstrations to idealized theoretical predictions of extremely large resistance contrasts, they collectively highlight the potential of altermagnetic metals for magnetoresistive junctions. Future progress will depend on converting these intrinsic electronic properties into robust and reproducible device responses through deterministic Néel-vector control, domain stabilization, interface engineering, and improved tolerance to disorder, finite temperature, finite bias, and inelastic scattering. Achieving these goals could provide a route toward stray-field-free, dense, and high-speed magnetoresistive devices.

## 7. Summary and Outlook

Magnetic tunnel junctions have become one of the most important platforms for exploring spin-dependent transport and developing spintronic technologies. The progress of MTJs has historically been closely associated with the optimization of magnetic electrodes, tunnel barriers, and their interfaces. In this Perspective, we have highlighted that the spin polarization relevant to tunneling transport is not determined solely by the intrinsic magnetic moment or bulk electronic structure of an electrode. Instead, it emerges from the interplay among magnetic order, electronic states, crystal symmetry, orbital character, and interface-dependent transmission. This viewpoint provides a common framework for understanding the diverse mechanisms underlying spin-polarized tunneling in ferromagnetic, antiferromagnetic, and altermagnetic electrode systems.

Different electrode classes have achieved different levels of technological maturity and face distinct challenges. Ferromagnetic electrodes, particularly CoFeB-based systems, remain the most established platform for practical MTJs due to their high tunnel magnetoresistance, reliable switching, and compatibility with industrial fabrication. Future progress will mainly rely on further improving interface quality, thermal stability, and scaling performance, while addressing issues associated with stray fields and energy consumption. Antiferromagnetic electrodes offer attractive properties including negligible net magnetization and robustness against external magnetic fields, but their practical development is still limited by challenges in controlling and electrically manipulating the Néel vector, managing magnetic domains, and achieving reproducible interfaces with well-defined transport characteristics. Altermagnetic electrodes represent an emerging opportunity to generate spin-polarized transport from symmetry-driven momentum-dependent spin splitting without macroscopic magnetization. However, experimental implementation remains at an early stage, and further efforts are needed to establish reliable material synthesis, domain control, interface engineering, and reproducible device demonstrations.

Looking ahead, the development of magnetic metallic electrodes will require a closer integration of materials design and device engineering. A central challenge is to bridge the gap between ideal material properties and realistic tunneling devices, where defects, interfaces, finite temperature, and electrical operation conditions strongly influence the final performance. Continued advances in interface characterization, atomic-scale fabrication, and electrical control of magnetic order will be essential for determining which emerging magnetic materials can ultimately complement or extend conventional ferromagnetic electrodes in future spintronic applications.

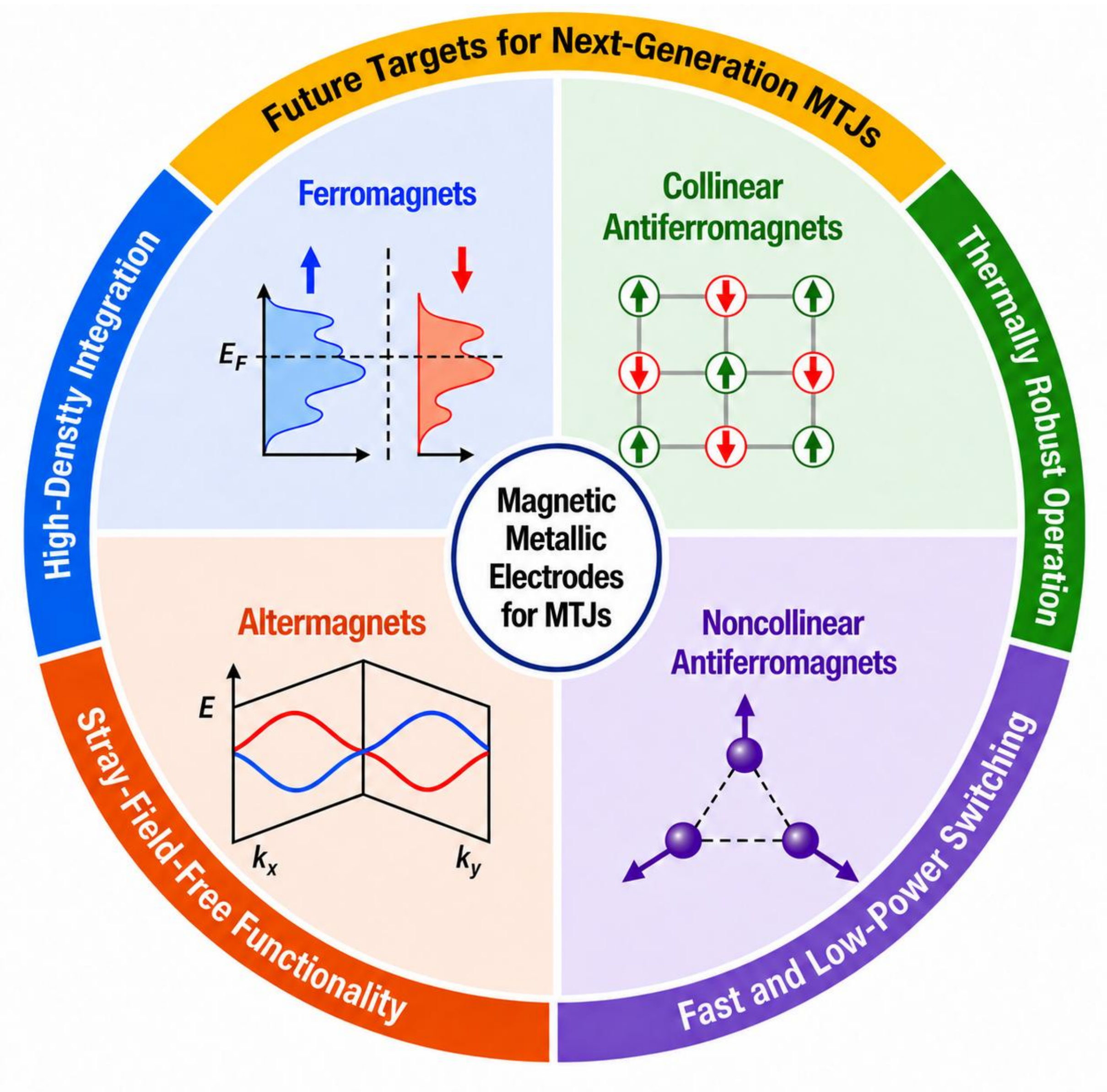

**Figure 1.** Schematic overview of ferromagnetic, collinear antiferromagnetic, noncollinear antiferromagnetic, and altermagnetic metallic electrodes for magnetic tunnel junctions (MTJs).

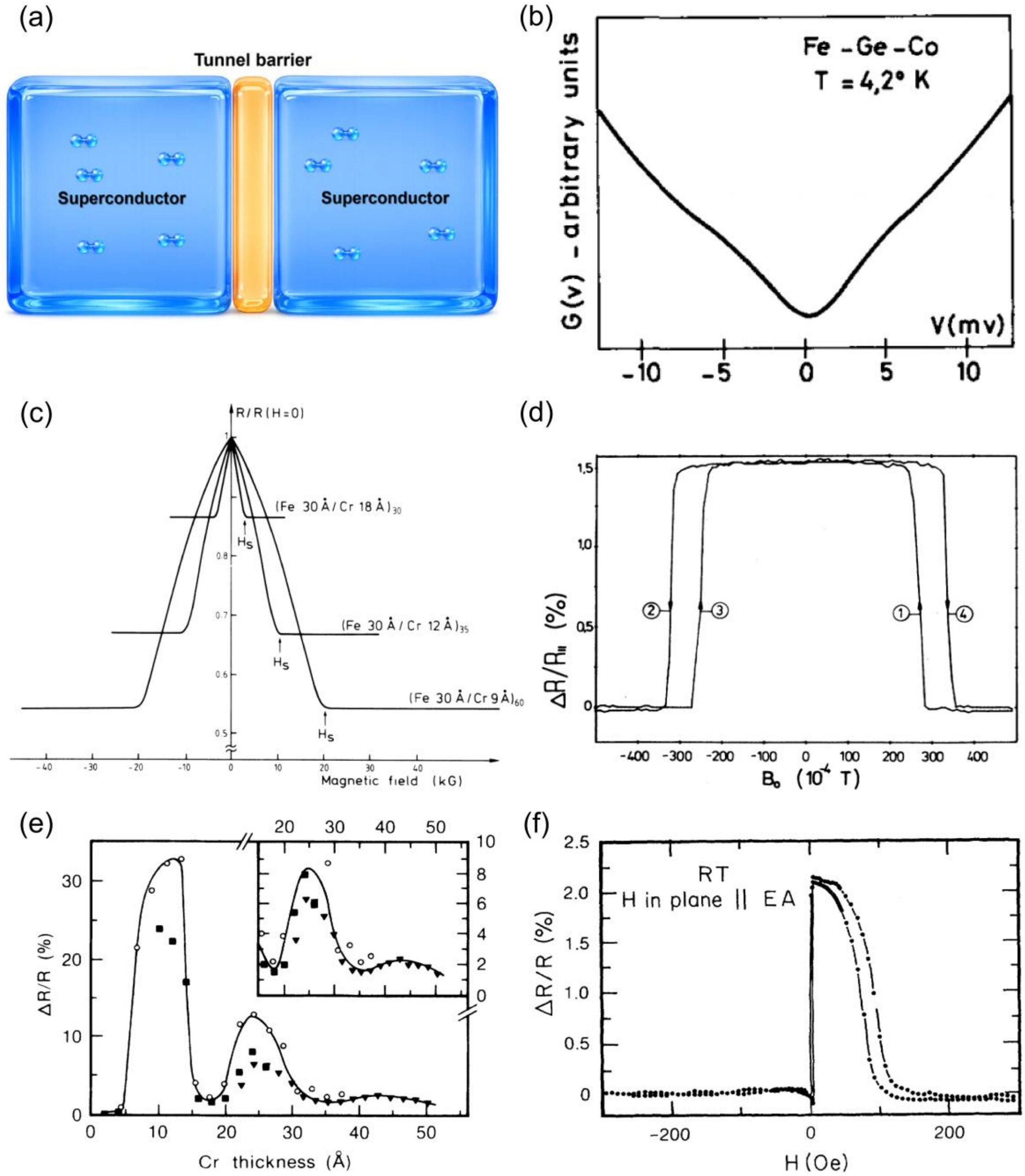

**Figure 2**. (a) Schematic illustration of a Josephson junction. (b) Bias-dependent conductance of the Fe/Ge/Co junction measured at 4.2 K [9]. Copyright 1975, Elsevier. (c) Magnetic-field dependence of the normalized resistance in Fe/Cr multilayers with different numbers of bilayer repeats [36]. Copyright 1988, American Physical Society. (d) Field-dependent magnetoresistance of an Fe/Cr/Fe trilayer [37]. Copyright 1989, American Physical Society. (e) Dependence of the magnetoresistance on Cr spacer thickness [40]. Copyright 1990, American Physical Society. (f) Room-temperature spin-valve response obtained by independently controlling the magnetic states of a soft free layer and an exchange-biased reference layer [41]. Copyright 1991, American Physical Society.

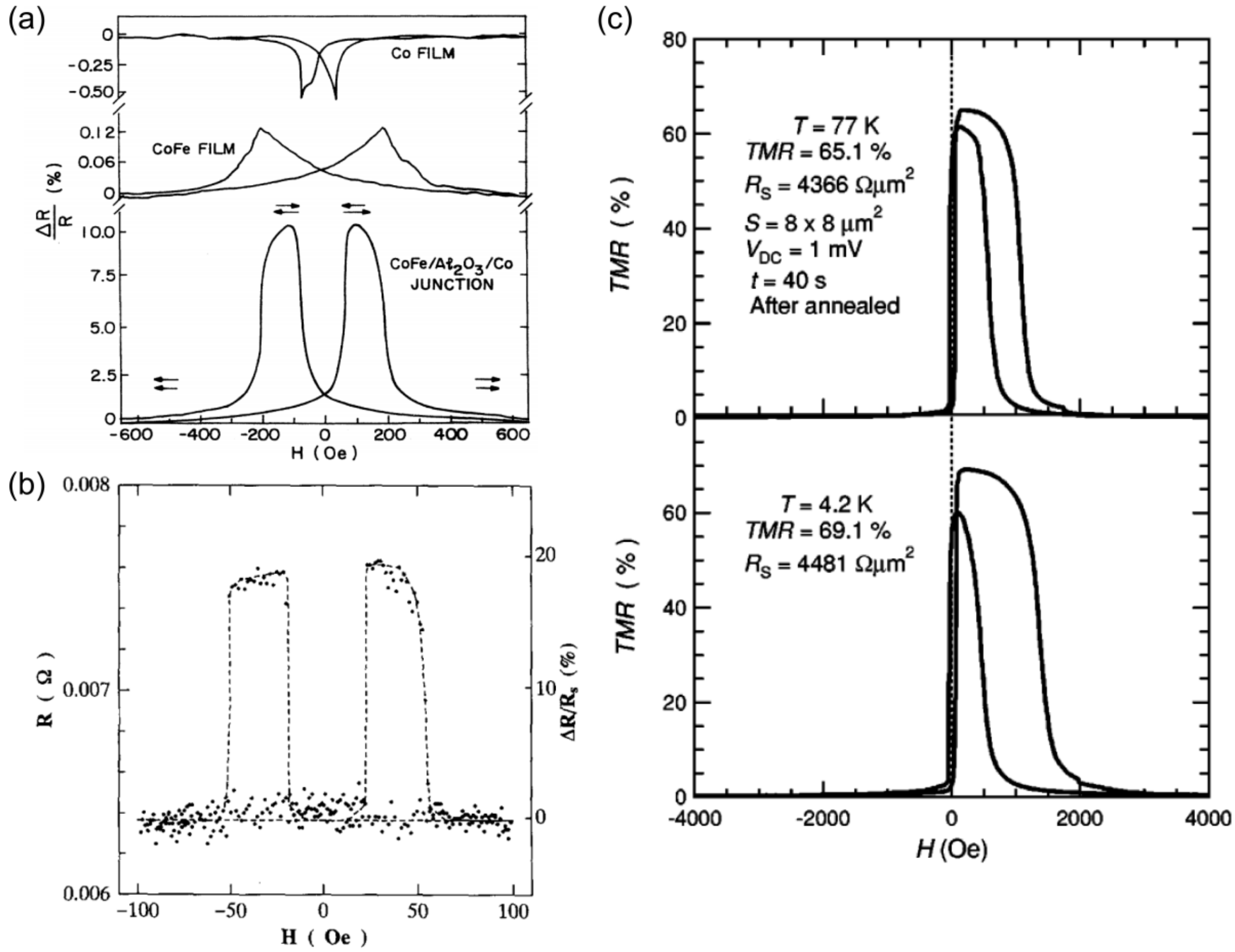

**Figure 3**. (a) Magnetoresistance of single Co and CoFe films compared with the much larger resistance change of a CoFe/$Al_2O_3$/Co tunnel junction [10]. Copyright 1995, American Physical Society. (b) Magnetic-field dependence of the resistance and the corresponding tunneling magnetoresistance (TMR) ratio in an Fe/$Al_2O_3$/Fe junction [11]. Copyright 1995, Elsevier. (c) TMR loops of an annealed $Co_{75}Fe_{25}$/$AlO_x$/$Co_{75}Fe_{25}$ junction measured at 77 K and 4.2 K [42]. Copyright 2000, AIP Publishing.

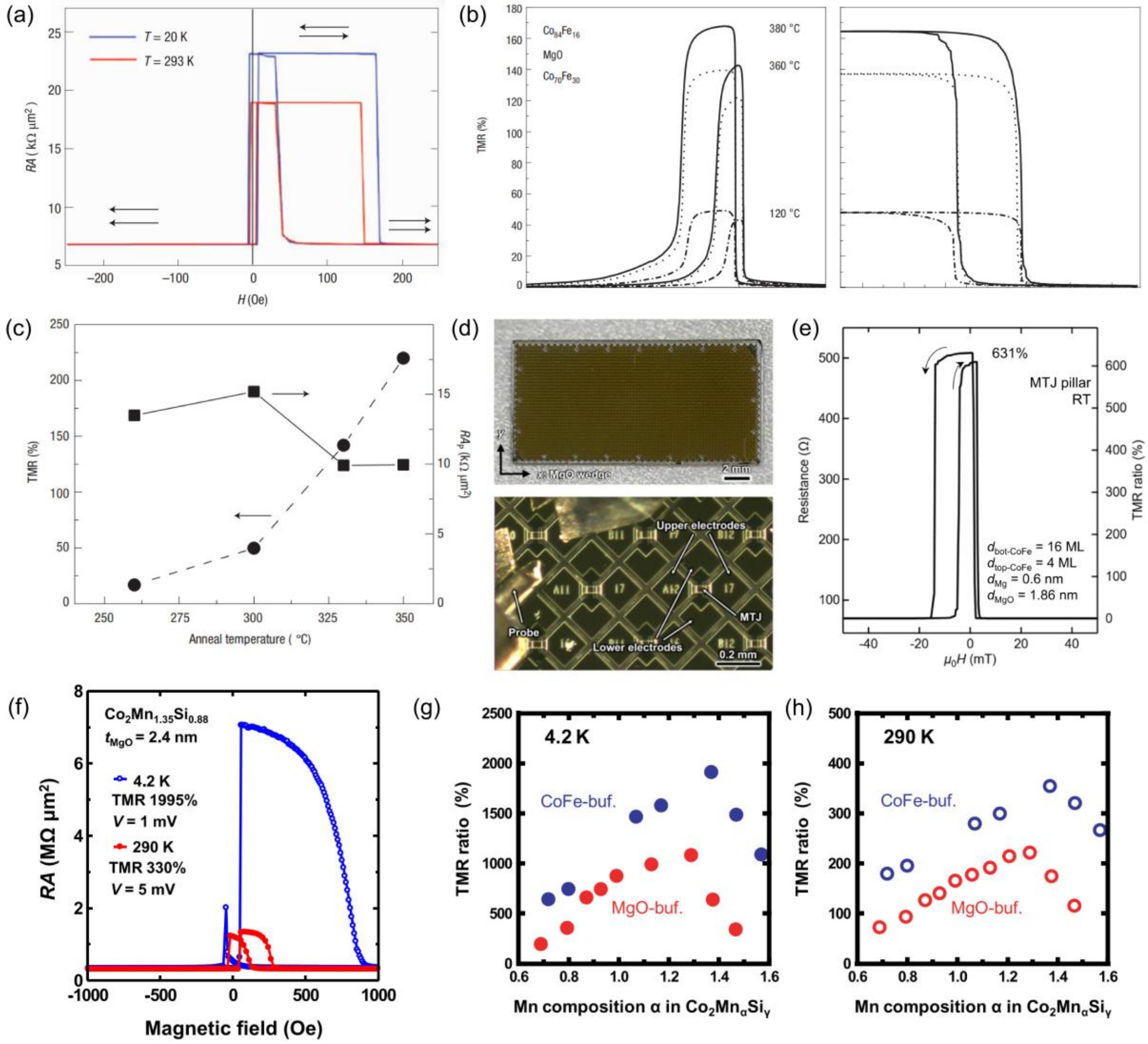

**Figure 4.** (a) Resistance-area product as a function of magnetic field for an epitaxial Fe/MgO/Fe(001) junction at 20 K and 293 K [13]. Copyright 2004, Springer Nature. (b,c) Representative TMR loops and the annealing-temperature dependence of TMR and the parallel-state resistance-area product [14]. Copyright 2004, Springer Nature. (d) Photograph of an MgO wedge sample and optical micrograph of patterned MTJ devices [47]. Copyright 2023, AIP Publishing. (e) Room-temperature resistance and TMR loop of an epitaxial CoFe/MgO/CoFe junction [47]. Copyright 2023, AIP Publishing. (f) Field-dependent resistance-area product of a $Co_2Mn_{1.35}Si_{0.88}$/MgO-based junction [57]. Copyright 2012, AIP Publishing. (g,h) TMR ratio versus Mn composition in $Co_2Mn_\alpha Si_\gamma$ electrodes at 4.2 K and 290 K for CoFe-buffered and MgO-buffered junctions [57]. Copyright 2012, AIP Publishing.

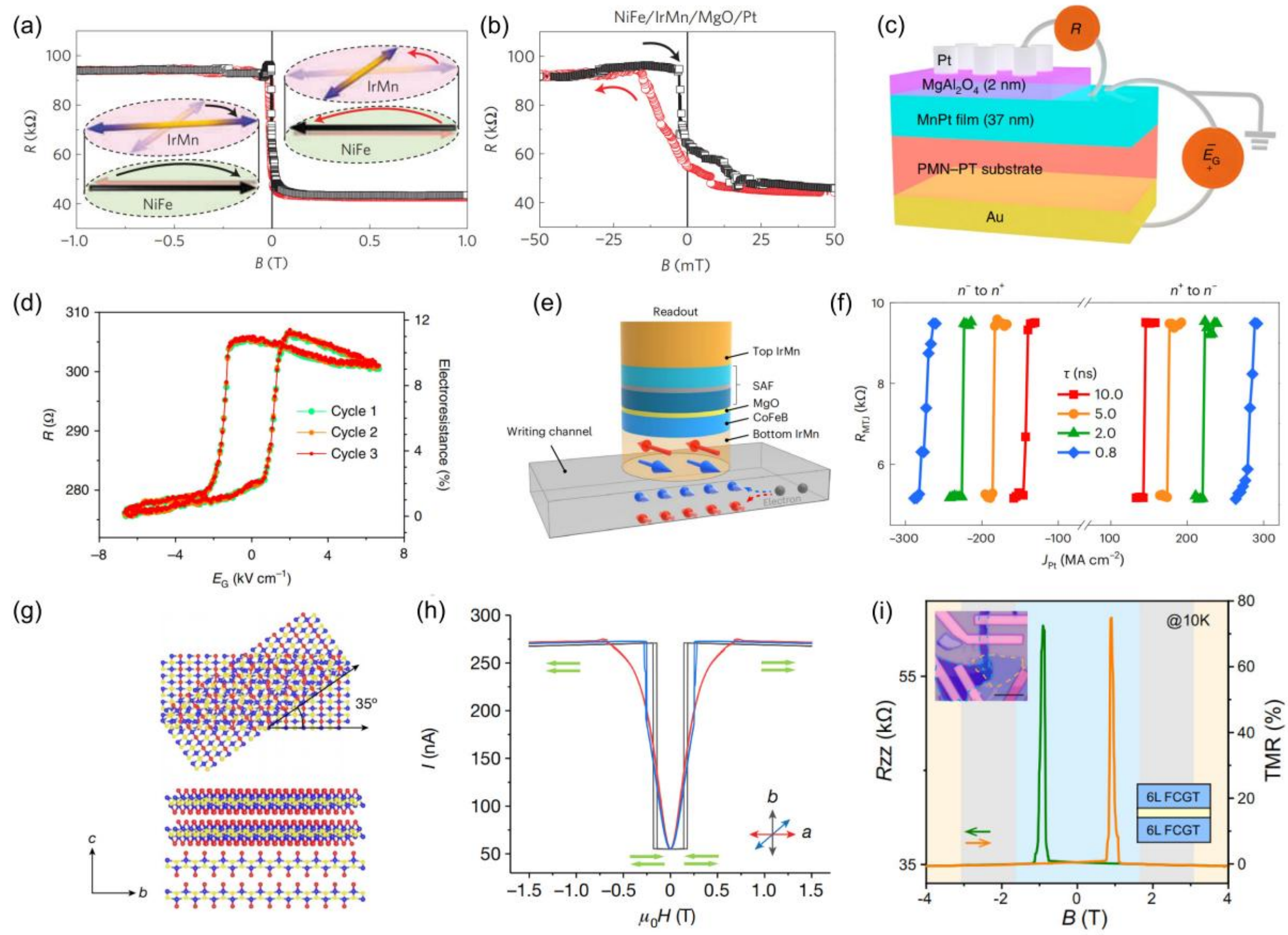

**Figure 5.** (a) Magnetic-field dependence of the resistance of a NiFe/IrMn/MgO/Pt junction over a wide field range [83]. Copyright 2011, Springer Nature. (b) Low-field resistance loop of the NiFe/IrMn/MgO/Pt junction [83]. Copyright 2011, Springer Nature. (c) Schematic of a Pt/$MgAl_2O_4$/MnPt tunnel junction fabricated on a piezoelectric PMN-PT substrate [86]. Copyright 2019, Springer Nature. (d) Junction resistance and electroresistance as functions of the applied electric field over three consecutive cycles [86]. Copyright 2019, Springer Nature. (e) Schematic of a three-terminal Pt/IrMn/CoFeB/MgO/CoFeB device with a lateral writing channel and a vertical TMR readout path [87]. Copyright 2023, Springer Nature. (f) MTJ resistance as a function of the writing-current density for pulse durations ranging from 0.8 to 10 ns [87]. Copyright 2023, Springer Nature. (g) Top and side views of the atomic structure of two CrSBr bilayers stacked with a twist angle of 35 degrees [89]. Copyright 2024, Springer Nature. (h) Representative field-dependent tunneling-current hysteresis of the twisted CrSBr junction for selected in-plane field orientations [89]. Copyright 2024, Springer Nature. (i) Out-of-plane resistance and TMR of an all-antiferromagnetic FCGT/$WSe_2$/FCGT junction measured as functions of magnetic field at 10 K; the inset shows an optical micrograph of the device [92]. Copyright 2025, Springer Nature.

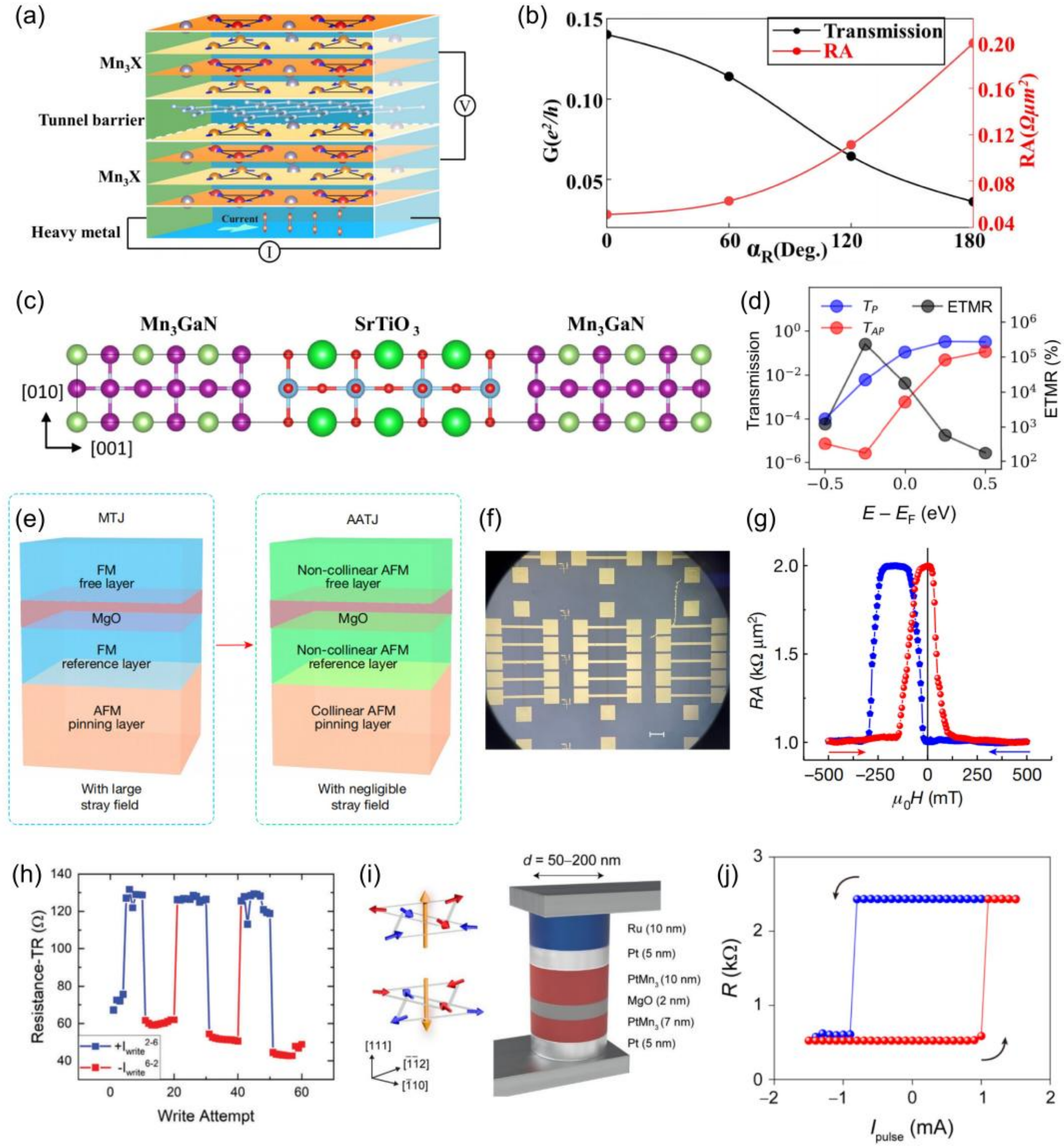

**Figure 6.** (a) Schematic of a $Mn_3X$/tunnel-barrier/$Mn_3X$ junction integrated with a heavy-metal writing layer [94]. Copyright 2022, American Physical Society. (b) Calculated evolution of transmission and resistance-area product with the relative rotation angle of the noncollinear magnetic orders [94]. Copyright 2022, American Physical Society. (c) Atomic structure of a $Mn_3GaN$/$SrTiO_3$/$Mn_3GaN$ junction [95]. Copyright 2024, Springer Nature. (d) Calculated parallel- and antiparallel-state transmission and the corresponding energy-dependent TMR [95]. Copyright 2024, Springer Nature. (e) Comparison between a conventional ferromagnetic MTJ and an all-antiferromagnetic tunnel junction containing noncollinear free and reference electrodes [109]. Copyright 2023, Springer Nature. (f) Optical micrograph of a fabricated $Mn_3Pt$-based junction array [109]. Copyright 2023, Springer Nature. (g) Field-dependent resistance-area product showing room-temperature TMR in an $Mn_3Pt$/MgO/$Mn_3Pt$ junction [109]. Copyright 2023, Springer Nature. (h) Reproducible electrical switching between resistance states using alternating write-current polarities in a silicon-compatible three-terminal device [111]. Copyright 2024, John Wiley and Sons. (i) Layer structure and magnetic configurations of a nanoscale $Mn_3Pt$/MgO/$Mn_3Pt$ junction [112]. Copyright 2025. (j) Hysteretic resistance switching driven by vertical current pulses [112]. Copyright 2025.

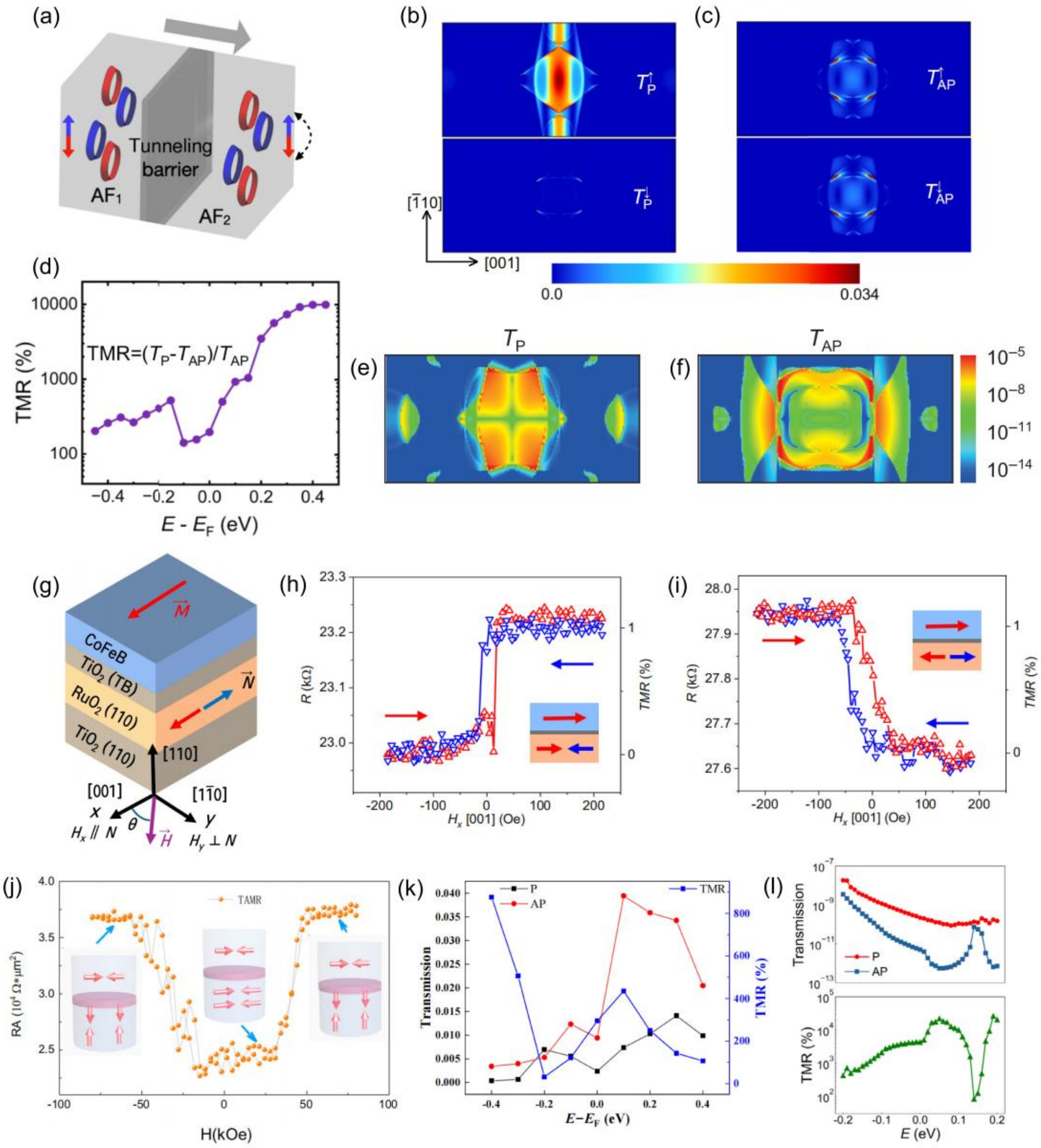

**Figure 7.** (a) Schematic of an antiferromagnetic tunnel junction in which the conductance depends on the relative orientations of two compensated magnetic electrodes [19]. Copyright 2022, American Physical Society. (b,c) Momentum-resolved transmission distributions for parallel and antiparallel magnetic configurations [141]. Copyright 2023, American Physical Society. (d) Calculated TMR as a function of energy relative to the Fermi level [141]. Copyright 2023, American Physical Society. (e,f) Momentum-resolved transmission in the parallel and antiparallel states for a facet-selected $RuO_2$-based junction [142]. Copyright 2024, American Physical Society. (g) Schematic of a hybrid $CoFeB/TiO_2/RuO_2/TiO_2$ junction and the orientations of the CoFeB magnetization and $RuO_2$ Néel vector [143]. Copyright 2025, American Physical Society. (h,i) Field-dependent resistance and TMR for two opposite $RuO_2$ Néel-vector states [143]. Copyright 2025, American Physical Society. (j) Resistance-area product of an epitaxial $RuO_2/MgO/RuO_2$ junction during a field-induced spin-flop transition between parallel and approximately orthogonal Néel-vector configurations [144]. Copyright 2025, Springer Nature. (k) Calculated parallel- and antiparallel-state transmissions and the corresponding TMR as functions of energy [145]. Copyright 2025, Institute of Physics. (l) Energy-dependent parallel and antiparallel transmissions plotted on a logarithmic scale, together with the resulting TMR [149]. Copyright 2026, John Wiley and Sons.